\documentstyle [12pt] {article}
\begin{document}
\def\be{\begin{eqnarray}}
\def\en{\end{eqnarray}}
\def\up{\uparrow}
\def\dw{\downarrow}
\def\non{\nonumber}
\def\la{\langle}
\def\ra{\rangle}
\def\ep{\varepsilon}
\def\bc{{\Lambda_b\Lambda_c}}
\def\cs{{\Lambda_c\Lambda}}
\def\bs{{\Lambda_b\Lambda}}
\def\b{\Lambda_b\to J/\psi\Lambda}
\def\j{{J/\psi}}
\def\pr{{\sl Phys. Rev.}~}
\def\prl{{\sl Phys. Rev. Lett.}~}
\def\pl{{\sl Phys. Lett.}~}
\def\np{{\sl Nucl. Phys.}~}
\def\zp{{\sl Z. Phys.}~}

\font\el=cmbx10 scaled \magstep2
{\obeylines
\hfill IP-ASTP-03-95
\hfill June, 1995
\hfill (Revised)}

\vskip 1.5 cm

\centerline{\large\bf 1/M Corrections to Baryonic Form Factors in the Quark
Model}
\medskip
\bigskip
\medskip
\centerline{\bf Hai-Yang Cheng and B. Tseng}
\medskip
\centerline{ Institute of Physics, Academia Sinica}
\centerline{Taipei, Taiwan 11529, Republic of China}
\bigskip
\bigskip
\bigskip
\centerline{\bf Abstract}
\bigskip
{\small
    Weak current-induced baryonic form factors at zero recoil are evaluated
in the rest frame of the heavy parent baryon using the nonrelativistic quark
model. Contrary to previous similar work in the literature, our quark model
results do satisfy the constraints imposed by heavy quark symmetry
for heavy-heavy baryon transitions at the symmetric point $v\cdot v'=1$
and are in agreement with the predictions of the heavy quark effective
theory for antitriplet-antitriplet heavy baryon form factors at zero recoil
evaluated to order $1/m_Q$. Furthermore,
the quark model approach has the merit that it
is applicable to any heavy-heavy and heavy-light baryonic transitions at
maximum $q^2$. Assuming a dipole $q^2$ behavior, we have applied the quark
model form factors to nonleptonic, semileptonic
and weak radiative decays of the heavy baryons. It is emphasized that the
flavor suppression factor occurring in many heavy-light baryonic transitions,
which is unfortunately overlooked in most literature,
is very crucial towards
an agreement between theory and experiment for the semileptonic decay
$\Lambda_c\to\Lambda e^+\nu_e$. Predictions for the decay modes $\b,~\Lambda_c
\to p\phi,~\Lambda_b\to\Lambda\gamma$, $\Xi_b\to\Xi\gamma$, and for the
semileptonic decays of $\Lambda_b,~\Xi_{b,c}$ and $\Omega_b$ are presented.

}

\pagebreak

\noindent{\bf I.~~Introduction}
\vskip 0.15cm
   In the heavy quark effective theory (HQET), there are two different types
of $1/m_Q$ corrections to the hadronic form factors: one from the
$1/m_Q$ correction to the current operators, and the other from the
presence of higher dimensional operators in the effective Lagrangian [1].
The latter amounts to the hadronic wave-function modifications. In general,
the predictive power of HQET for $1/m_Q$ effects is very limited by the fact
that we do not know how to carry out first-principles calculations for the
hadronic matrix elements in which higher dimensional kinetic and
chromo-magnetic operators $O_1$ and $O_2$ are inserted. Consequently,
several new unknown functions are necessarily introduced besides the leading
Isgur-Wise functions. For example, to order $\Lambda_{\rm QCD}/m_c$, there are
four new subleading Isgur-Wise functions $\eta(\omega),~\chi_1(\omega),~
\chi_2(\omega)$ and $\chi_3(\omega)$
\footnote{We follow the notation of Ref.[1] for subleading Isgur-Wise
functions. The new function $\eta(\omega)$ arises
from the matrix element of the $1/m_Q$-expanded current operator.}
for $B\to D$ transition, whose
normalizations are not determined except that $\chi_1$ and
$\chi_3$ vanish at the zero-recoil point $\omega\equiv v\cdot v'=1$ [2]. Since
the Isgur-Wise functions are not calculable from perturbative QCD or HQET,
a calculation of them should be resorted to some models.
It is known that the Isgur-Wise
functions have some simple expressions in the quark model. Denoting the
heavy meson wave function by
\be
\psi=\psi_0+\psi_{\rm kin}+\psi_{\rm mag}+\cdots,
\en
where $\psi_0$ is the wave function in the heavy quark limit,
$\psi_{\rm kin}$ and $\psi_{\rm mag}$ are the $1/m_Q$ corrections to
the wave function due to the operators $O_1$ and $O_2$
respectively, the Isgur-Wise function $\xi(v\cdot v')$ simply measures
the degree of overlap between the wave functions $\psi_0(v)$ and
$\psi_0(v')$, while $\chi_1$ ($\chi_3$) can be expressed as the overlap
integral of $\psi_{\rm kin}$ ($\psi_{\rm mag}$) and $\psi_0$ [3].

    In the heavy baryon case, there exist three baryonic Isgur-Wise functions
in the
heavy quark limit: $\zeta(\omega)$ for antitriplet-antitriplet transition,
and $\xi_1(\omega),~\xi_2(\omega)$ for sextet-sextet transition. In principle,
these functions are also calculable in the quark model though they are more
complicated. However, a tremendous simplification occurs in the
antitriplet-antitriplet heavy baryon transition, e.g. $\Lambda_b\to
\Lambda_c$: $1/m_Q$ corrections only amount to renormalizing the function
$\zeta(\omega)$ and no further new function is needed [4]. This
simplification stems from the fact that the chromo-magnetic operator does not
contribute to $\Lambda_b\to \Lambda_c$ and that the diquark of the antitriplet
heavy baryon is a spin singlet. Therefore, $1/m_b$ and $1/m_c$ corrections
to $\Lambda_b\to\Lambda_c$ and $\Xi_b\to\Xi_c$
form factors are predictable in HQET and
certain heavy quark symmetry relations among baryonic form factors remain
intact. Since HQET is a theory, its prediction is model independent.

   Going beyond the antitriplet-antitriplet heavy baryon transition, the
predictive power of HQET for form factors at order $1/m_Q$ is lost owing to
the fact that $1/m_Q$ corrections due to wave function modifications arising
from $O_1$ and especially $O_2$ are not calculable by perturbative QCD.
Therefore, it is appealing to have model calculations which enable us to
estimate the $1/m_Q$ corrections for other baryon form factors.
To our knowledge, two different quark-model calculations [5,6] are available
in the literature. In order to ensure that quark model results are reliable
and trustworthy, model predictions, when applied to heavy-heavy baryon
transitions, must satisfy all the constraints imposed by heavy quark symmetry.
In the heavy quark limit, normalization of the form factors at zero recoil is
fixed by heavy quark symmetry. To order $1/m_Q$, antitriplet-antitriplet
(i.e. $\Lambda_Q\to\Lambda_{Q'}$, $\Xi_Q\to\Xi_{Q'}$) form factors are also
calculable in HQET. Unfortunately, none of the calculations presented
in [5,6] is in agreement with the predictions of HQET. For example,
several heavy quark symmetry relations between baryon form factors are
not respected in Ref.[5]. While this discrepancy is
resolved in Ref.[6], the $1/m_Q$ corrections obtained in this reference are
still inconsistent with HQET in magnitude.

     The purpose of the present paper is to demonstrate that the
nonrelativistic quark model for baryon form factors, when evaluated properly,
does respect all the heavy quark symmetry constraints, including $1/m_Q$
corrections to $\Lambda_Q\to\Lambda_{Q'}$ ($\Xi_Q\to\Xi_{Q'}$) form factors
predicted by HQET. As a consequence, this model does incorporate the features
of heavy quark symmetry and can be used to compute form factors beyond
the arena of HQET. Since the quark-model wave function best resembles the
hadronic state in the rest frame, we will thus first evaluate form factors at
the zero recoil kinetic point. Instead of evaluating the baryonic Isgur-Wise
functions, which are beyond the scope of a nonrelativistic quark model,
we will make the conventional pole dominance assumption for the $q^2$
dependence to extrapolate the form factors from maximum $q^2$ to the desired
$q^2$ point. Since corrections to the form factors due to the modified wave
functions vanish at zero recoil (see Sec. II),
the nonrelativistic quark model applies equally well to the
sextet-sextet heavy baryon transition, e.g. $\Omega_b\to\Omega_c$ at the
symmetric point $v\cdot v'=1$. Moreover, it becomes meaningful to consider
in this model the $1/m_s$ corrections to, for example, $\Lambda_Q\to\Lambda$
and $\Xi_Q\to\Xi$ form factors at maximum $q^2$ so long as the recoil
momentum is smaller than the $m_s$ scale.

   The layout of the present paper is as follows. In Sec. II we will derive,
within the framework of the nonrelativistic quark model, the $1/m_Q$
and $1/m_{q}$ corrections (a distinction between $m_Q$ and $m_q$ will be
defined in Sec. II), coming from the current operator,
to the baryonic form factors at zero recoil.
The HQET predictions for $\Lambda_b\to\Lambda_c$ at order $1/m_Q$
are reproduced in this quark-model calculation. Assuming a pole
behavior for the $q^2$ dependence of the form factors, we will apply in
Sec. III the
quark-model results for baryonic form factors to nonleptonic weak decays,
semileptonic decays and weak radiative decays. Sec. IV comes to our
discussion and conclusion.

\vskip 0.70 cm
\noindent{\bf II.~~Baryonic Form Factors in the Nonrelativistic Quark Model}
\vskip 0.20 cm
     The general expression for the baryonic transition $B_i\to B_f$ reads
\be
\la B_f(p_f)|V_\mu-A_\mu|B_i(p_i)\ra &=& \bar{u}_f
[f_1(q^2)\gamma_\mu+if_2(q^2)\sigma_{\mu\nu}q^\nu+f_3(q^2)q_\mu   \non \\
&& -(g_1(q^2)\gamma_\mu+ig_2(q^2)\sigma_{\mu\nu}q^\nu+g_3(q^2)q_\mu)\gamma_5]
u_i,
\en
where $q=p_i-p_f$. When both baryons are heavy, it is also convenient to
parametrize the matrix element in terms of the velocities $v$ and $v'$:
\be
\la B_f(v')|V_\mu-A_\mu|B_i(v)\ra &=& \bar{u}_f
[F_1(\omega)\gamma_\mu+F_2(\omega)v_\mu+F_3(\omega)v'_\mu   \non \\
&& -(G_1(\omega)\gamma_\mu+G_2(\omega)v_\mu+G_3(\omega)v'_\mu)\gamma_5]u_i,
\en
with $\omega\equiv v\cdot v'$.  The form factors $F_i$ and $G_i$ are related
to $f_i$ and $g_i$ via
\be
f_1 &=& F_1+{1\over 2}(m_i+m_f)\left({F_2\over m_i}+{F_3\over m_f}\right),
\non \\
f_2 &=& {1\over 2}\left({F_2\over m_i}+{F_3\over m_f}\right), \non \\
f_3 &=& {1\over 2}\left({F_2\over m_i}-{F_3\over m_f}\right),  \\
g_1 &=& G_1-{1\over 2}(m_i-m_f)\left({G_2\over m_i}+{G_3\over m_f}\right),
\non \\
g_2 &=& {1\over 2}\left({G_2\over m_i}+{G_3\over m_f}\right), \non \\
g_3 &=& {1\over 2}\left({G_2\over m_i}-{G_3\over m_f}\right), \non
\en
where $m_i$ ($m_f$) is the mass of $B_i~(B_f)$.
Since the quark model is most trustworthy when the baryon is static, we will
thus evaluate the form factors at zero recoil $\vec{q}=0$ (or $q^2=(m_i-m_f)
^2$) in the rest frame of the parent baryon $B_i$. Note that in order to
determine the form factors $f_{2,3}$ and $g_{2,3}$, we need to keep the
small recoil momentum $\vec{q}$ in Eq.(2) when recasting the 4-component
Dirac spinors in terms of the 2-component Pauli spinors.

     In order to calculate the form factors using the nonrelativistic
quark model, we write [5]
\be
\la B_f|V_0|B_i\ra &=& \chi_f^\dagger\tilde{V}_0(q^2)\chi_i,  \non \\
\la B_f|A_0|B_i\ra &=& \chi_f^\dagger\vec{\sigma}\cdot\vec{q}\tilde{A}_0(q^2)
\chi_i,  \non \\
\la B_f|\vec{V}|B_i\ra &=& \chi_f^\dagger[\vec{q}\,\tilde{V}_V(q^2)+i
\vec{\sigma}\times\vec{q}\,\tilde{V}_M(q^2)]\chi_i,   \\
\la B_f|\vec{A}|B_i\ra &=& \chi_f^\dagger[\vec{\sigma}\tilde{A}_S(q^2)
+\vec{q}\,(\vec{\sigma}\cdot\vec{q})\tilde{A}_T(q^2)]\chi_i,  \non
\en
where $\chi$ is a Pauli spinor. In the rest frame of $B_i$, we find from
Eqs.(2) and (5) that the scalar coefficients $\tilde{V}
$ and $\tilde{A}$ at maximum $q^2$ are given by
\be
\tilde{V}_0(q^2_m) &=& f_1+\Delta mf_3,   \non \\
\tilde{V}_V(q^2_m) &=& {1\over 2m_f}(-f_1+\Delta mf_2)+f_3,   \non \\
\tilde{V}_M(q^2_m) &=& {1\over 2m_f}[-f_1+(m_i+m_f)f_2],   \non \\
\tilde{A}_0(q^2_m) &=& {1\over 2m_f}(-g_1+\Delta mg_3)+g_2,   \\
\tilde{A}_S(q^2_m) &=& g_1+\Delta mg_2,   \non \\
\tilde{A}_T(q^2_m) &=& {1\over 2m_f}(-g_2+g_3),   \non
\en
where $q_m^2\equiv q_{\rm max}^2=(\Delta m)^2$ and $\Delta m=m_i-m_f$.
Inverting the above equations gives
\be
f_1(q_m^2) &=& \left(1-{\Delta m\over 2m_i}\right)\tilde{V}_0-{\Delta m(m_i
+m_f)\over 2m_i}\tilde{V}_V+{(\Delta m)^2\over 2m_i}\tilde{V}_M,  \non \\
f_2(q^2_m) &=& {1\over 2m_i}\tilde{V}_0-{\Delta m\over 2m_i}\tilde{V}_V+
{m_i+m_f\over 2m_i}\tilde{V}_M,   \non  \\
f_3(q^2_m) &=& {1\over 2m_i}\tilde{V}_0+{m_i+m_f\over 2m_i}\tilde{V}_V-
{\Delta m\over 2m_i}\tilde{V}_M,     \\
g_1(q_m^2) &=& \left(1-{\Delta m\over 2m_i}\right)\tilde{A}_S-{m_f\Delta m
\over m_i}\tilde{A}_0+{m_f(\Delta m)^2\over m_i}\tilde{A}_T,  \non \\
g_2(q^2_m) &=& {1\over 2m_i}\tilde{A}_S+{m_f\over m_i}\tilde{A}_0-
{m_f\Delta m\over m_i}\tilde{A}_T,   \non  \\
g_3(q^2_m) &=& {1\over 2m_i}\tilde{A}_S+{m_f\over m_i}\tilde{A}_0+
{m_f(m_i+m_f)\over m_i}\tilde{A}_T.   \non
\en

Our next task is to employ the nonrelativistic quark model to evaluate
the coefficients $\tilde{V}$ and $\tilde{A}$ at $q^2=q^2_m$. We will follow
closely Ref.[6] for this task. Suppose that the parent baryon $B_i$
contains a heavy quark $Q$ and two light quarks $q_1$ and $q_2$ behavioring
as a spectator diquark, and that the final baryon $B_f$ is composed of
the quark $q$ (being a heavy quark $Q'$ or a $s$ quark)
and the same light diquark as in $B_i$. Denoting the spatial
coordinates of the three quarks in $B_i$ by $\vec{r}_Q,~\vec{r}_1$ and
$\vec{r}_2$, we define the relative coordinates
\be
\vec{R}=\,{\sum m_j\vec{r}_j\over \tilde{m}_i},~~~~~\vec{r}_{12}=\vec{r}_2-
\vec{r}_1,~~~~~\vec{r}_\ell={m_1\vec{r}_1+m_2\vec{r}_2\over m_1+m_2}-\vec{r}_Q,
\en
where $\tilde{m}_i=m_Q+m_1+m_2$, which is in practice close to $m_i$,
so that $\vec{r}_{12}$ is the relative coordinate of the two light quarks, and
$\vec{r}_\ell$ is the relative coordinate of $Q$ and the c.m. of the diquark.
It is easily shown that the corresponding relative momenta are [6]
\be
\vec{P} &=& \vec{p}_1+\vec{p}_2+\vec{p}_3,  \non  \\
\vec{p}_{12} &=& {m_1\over m_1+m_2}\vec{p}_2-{m_2\over m_1+m_2}\vec{p}_1, \\
\vec{\ell} &=& {m_Q\over \tilde{m}_i}(\vec{p}_1+\vec{p}_2)-{m_1+m_2\over
\tilde{m}_i}\,\vec{p}_Q.   \non
\en
In the rest frame of the parent baryon, the momenta $Q$ and $q$ are
related to the relative momentum $\vec{\ell}$ via
\be
\vec{p}_Q=-\vec{\ell},~~~~~\vec{p}_{q}=-\vec{q}-\vec
{\ell},
\en
and the relative momenta of the quarks in the baryon $B_f$ denoted with
primes are related to that in $B_i$ by
\be
\vec{p'}_{12}=\,\vec{p}_{12},~~~~~\vec{\ell}'=\vec{\ell}+{m_1+m_2\over
\tilde{m}_f}\,\vec{q},
\en
with $\tilde{m}_f=m_q+m_1+m_2$.
Note that the recoil momentum of the daughter baryon $B_f$ is $-\vec{q}$.

     The baryon state is represented in the nonrelativistic quark model by
\be
|B_i(\vec{P},s)\ra =\int d^3\vec{p}_{12}d^3\vec{\ell}\,\phi(\vec{p}_{12},
\vec{\ell}\,)\sum_{s_1,s_2,s_Q}C^s_{s_1,s_2,s_Q}|Q(\vec{p}_Q,s_Q),
q_1(\vec{p}_1,s_1),q_2(\vec{p}_2,s_2)\ra,
\en
where $C^s_{s_1,s_2,s_Q}$ is the Clebsch-Gordan coefficient for the
combination of three constituent quarks into a spin-${1\over 2}$ baryon
with the spin component $s$ along the $z$ direction, and $\phi(\vec{p}_{12},
\vec{\ell}\,)$ is the momentum wave function satisfying the normalization
condition:
\be
\int d^3\vec{p}_{12}d^3\vec{l}~|\phi(\vec{p}_{12},\vec{\ell}\,)|^2=1.
\en
We shall see that the form factors to be evaluated at zero recoil do not
depend on the explicit detail of $\phi(\vec{p}_{12},\vec{\ell}\,)$.
Consider the weak current $J_\mu=\bar{q}\gamma_\mu(1-\gamma_5)Q$. In the
quark model the hadronic matrix element in (2) becomes
\be
&& \la B_f(\vec{P'}=-\vec{q},s')|J_\mu|B_i(\vec{P}=0,s)\ra=
\int d^3\vec{p'}_{12}d^3\vec{\ell}'d^3\vec{p}_{12}d^3\vec{\ell}   \non \\
&  \times &\delta^3(
\vec{p'}_{12}-\vec{p}_{12})\delta^3\left(\vec{\ell}'-\vec{\ell}-{m_1+m_2\over
\tilde{m}_f}\vec{q}\right)\phi_f(\vec{p'}_{12},\vec{\ell}')\phi_i(\vec{p}_{12},
\vec{\ell}\,)\la s'|J_\mu|s\ra,
\en
with
\be
\la s'|J_\mu |s\ra=\sum C^{s'}_{s_1,s_2,s_q}C^s_{s_1,s_2,s_Q}\bar{q}(\vec{p}
_q,s_q)\gamma_\mu(1-\gamma_5)Q(\vec{p}_Q,s_Q).
\en
It will become clear shortly that it makes difference to choose
$(\vec{p'}_{12},\vec{\ell}')$ or $(\vec{p}_{12},\vec{\ell}\,)$ as the
integration variables after integrating over the $\delta$-functions.
We thus take the average
\footnote{The analogous expression obtained by Singleton [6] is
\be
\la B_f(-\vec{q},s')|J_\mu|B_i(\vec{0},s)\ra= \left({m_q\over\tilde{m}_f}
\right)^3 \int d^3\vec{p}_{12}
d^3\vec{\ell}~\phi_f(\vec{p'}_{12},\vec{\ell}')\phi_i(\vec{p}_{12},
\vec{\ell}\,)\la s'|J_\mu|s\ra.   \non
\en
He noticed that there would be a factor of
$(m_Q/\tilde{m}_i)^3$ instead if $(\vec{p'}_{12},\vec{\ell}')$ were
integrated over [these factors do not appear in our Eq.(14) or (16)]. It is
argued in [6] that these factors can be neglected since in the spectator model
the diquark mass should not affect the rate.}
\be
\la B_f(-\vec{q},s')|J_\mu|B_i(\vec{0},s)\ra=
 {1\over 2}\left[\int d^3\vec{p'}_{12}d^3\vec{\ell}'+\int d^3\vec{p}_{12}
d^3\vec{\ell}\,\right]\phi_f(\vec{p'}_{12},\vec{\ell}')\phi_i(\vec{p}_{12},
\vec{\ell}\,)\la s'|J_\mu|s\ra.
\en

  In the nonrelativistic limit, the Dirac spinors in (15) read
\be
\bar{q}(\vec{p}_q,s_q)=\,\chi^\dagger\left(1,~~\,-{\vec{\sigma}\cdot\vec{p}_q
\over 2m_q}\right),~~~Q(\vec{p}_Q,s_Q)=\left(\matrix{1   \cr {\vec{\sigma}
\cdot\vec{p}_{_Q}\over 2m_Q}   \cr}\right)\chi.
\en
Note that $\vec{p}_q=-(\vec{q}+\vec{\ell}),~\vec{p}_Q=-\vec{\ell}$ when $(
\vec{p}_{12},\vec{\ell}\,)$ are chosen to be the integration variables, and
$\vec{p}_q=-(m_q/ \tilde{m}_f)\vec{q}-\vec{\ell}',~\vec{p}_Q=\vec{q}(m_1+m_2)/
\tilde{m}_f-\vec{\ell}'$ for the integration variables $(\vec{p'}_{12},
\vec{\ell}')$. Obviously, the integration over $(\vec{p}_{12},\vec{\ell}\,)$
is in general different from that over $(\vec{p'}_{12},\vec{\ell'})$.

    Substituting (17) into (16) and noting that terms linear in $\vec{\ell}$
($\vec{\ell}')$ make no contribution after integrating over $\vec{\ell}$
$(\vec{\ell}')$, we find after some manipulation that the scalar
coefficients $\tilde{V}$ and $\tilde{A}$ evaluated at $\vec{q}=0$ are
\be
\tilde{V}_0(q^2_m)/N_{fi} &=& 1,   \non \\
\tilde{V}_V(q^2_m)/N_{fi} &=& -{1\over 2m_q}\left(1-{\bar{\Lambda}\over
2m_f}\right)+{\bar{\Lambda}\over 4m_fm_Q},   \non \\
\tilde{V}_M(q^2_m)/N'_{fi} &=& -{1\over 2m_q}\left(1-{\bar{\Lambda}\over
2m_f}\right)-{\bar{\Lambda}\over 4m_fm_Q},    \\
\tilde{A}_0(q^2_m)/N'_{fi} &=& -{1\over 2m_q}\left(1-{\bar{\Lambda}\over
2m_f}\right)+{\bar{\Lambda}\over 4m_fm_Q},   \non \\
\tilde{A}_S(q^2_m)/N'_{fi} &=& 1,  \non  \\
\tilde{A}_T(q^2_m)/N'_{fi} &=& -{\bar{\Lambda}\over 4m_f^2m_Q},  \non
\en
where use of the approximation $\tilde{m}_f\approx m_f$ has been made,
$\bar{\Lambda}\equiv m_f-m_q$, and
\be
N_{fi}=\,_{\rm flavor-spin}\la B_f|b^\dagger_q b_Q|B_i\ra_{\rm flavor-spin},~~~
N'_{fi}=\,_{\rm flavor-spin}\la B_f|b^\dagger_q b_Q\sigma_z^Q|B_i\ra_{\rm
flavor-spin},
\en
with $\sigma_Q$ acting on the heavy quark $Q$.
In deriving Eq.(18) we have applied the normalization condition (13) for
the momentum wave function by assuming flavor independence, $\phi_f=\phi_i$.
Since
\be
\la\chi_s|\sigma_z^3|\chi_s\ra=-{1\over 3},~~~~\la\chi_A|\sigma_z^3|\chi_A
\ra=\,1,
\en
where $\chi_s=(2\up\up\dw-\up\dw\up-\dw\up\up)/\sqrt{6}$ is the spin wave
function for the sextet heavy baryon and $\chi_A=(\up\dw\up-\dw\up\up)/
\sqrt{2}$ for the antitriplet heavy baryon, it is clear that [6]
\be
\eta\equiv {N'_{fi}\over N_{fi}}=\cases{ 1&for~antitriplet~baryon~$B_i$,  \cr
-{1\over 3}& for sextet~baryon~$B_i$.   \cr}
\en

   It follows from Eqs.(18) and (7) that the form factors at zero recoil are
given by
\be
f_1(q^2_m)/N_{fi} &=& 1-{\Delta m\over 2m_i}+{\Delta m\over
4m_im_q}\left(1-{\bar{\Lambda}\over 2m_f}\right)(m_i+m_f-\eta\Delta m)\non \\
&& -{\Delta m\over 8m_im_f}\,{\bar{\Lambda}\over m_Q}(m_i+m_f+\eta\Delta m),
\non  \\
f_2(q^2_m)/N_{fi} &=& {1\over 2m_i}+{1\over 4m_im_q}\left(1-{\bar{\Lambda}
\over 2m_f}\right)[\Delta m-(m_i+m_f)\eta]  \non \\
&& -{\bar{\Lambda}\over 8m_im_fm_Q}[\Delta m+(m_i+m_f)\eta],  \non \\
f_3(q^2_m)/N_{fi} &=& {1\over 2m_i}-{1\over 4m_im_q}\left(1-{\bar{\Lambda}
\over 2m_f}\right)(m_i+m_f-\eta\Delta m)   \non \\
&& +{\bar{\Lambda}\over 8m_im_fm_Q}(m_i+m_f+\eta\Delta m),  \\
g_1(q^2_m)/N_{fi} &=& \eta+{\Delta m\bar{\Lambda}\over 4}\left({1\over m_i
m_q}-{1\over m_fm_Q}\right)\eta,  \non \\
g_2(q^2_m)/N_{fi} &=& -{\bar{\Lambda}\over 4}\left({1\over m_i m_q}-
{1\over m_fm_Q}\right)\eta,  \non \\
g_3(q^2_m)/N_{fi} &=& -{\bar{\Lambda}\over 4}\left({1\over m_i m_q}+
{1\over m_fm_Q}\right)\eta,   \non
\en
with $\bar{\Lambda}=m_f-m_q$.
When both baryons are heavy, the form factors defined in Eq.(3) have the
following expressions at $\omega=1$:
\be
F_1(1)/N_{fi} &=& \left[1+{\bar{\Lambda}\over 2}\left({1\over m_q}+
{1\over m_Q}\right)\right]\eta,   \non \\
F_2(1)/N_{fi} &=& {1\over 2}(1-\eta)-{\bar{\Lambda}\over 2m_q}+
{\bar{\Lambda}\over 4}\left({1\over m_q}-{1\over m_Q}\right)(1-\eta), \non \\
F_3(1)/N_{fi} &=& {1\over 2}(1-\eta)-{\bar{\Lambda}\over 2m_Q}+
{\bar{\Lambda}\over 4}\left({1\over m_q}+{1\over m_Q}\right)(1-\eta), \\
G_1(1)/N_{fi} &=& \eta,~~~G_2(1)/N_{fi}=-{\bar{\Lambda}\over 2m_q}\eta,
{}~~~G_3(1)/N_{fi}={\bar{\Lambda}\over 2m_Q}\eta,   \non
\en
obtained from Eqs.(4) and (22). For antitriplet $\Lambda_b\to\Lambda_c$
or $\Xi_b\to\Xi_c$ transition, $N_{fi}=1$, $\eta=1$, so we have
\be
F_1^\bc(1) &=& 1+{\bar{\Lambda}\over 2}\left({1\over m_c}+{1\over m_b}\right),
{}~~~~G_1^\bc(1)=\,1\,,   \non \\
F_2^\bc(1) &=& G_2^\bc(1)=-{\bar{\Lambda}\over 2m_c},    \\
F_3^\bc(1) &=& -G_3^\bc(1)=-{\bar{\Lambda}\over 2m_b},    \non
\en
and
\be
f_1^\bc(q^2_m)=g_1^\bc(q^2_m) &=& 1+{\Delta m\bar{\Lambda}\over 4}\left({1
\over m_{\Lambda_b}m_c}-{1\over m_{\Lambda_c}m_b}\right),   \non \\
f_2^\bc(q^2_m)=g_3^\bc(q^2_m) &=& -{\bar{\Lambda}\over 4}\left({1\over m_{
\Lambda_b}m_c}+{1\over m_{\Lambda_c}m_b}\right),    \\
f_3^\bc(q^2_m)=g_2^\bc(q^2_m) &=& -{\bar{\Lambda}\over 4}\left({1\over m_{
\Lambda_b}m_c}-{1\over m_{\Lambda_c}m_b}\right),   \non
\en
and similar expressions for $\Xi_b\to\Xi_c$.
Therefore, there is only one independent $\Lambda_b\to\Lambda_c$ form factor
in the heavy quark limit.
The relevant HQET predictions to the zeroth order of $\alpha_s$ are [4]
\be
F_1^\bc(\omega) &=& 1+{\bar{\Lambda}\over 2}\left({1\over m_c}+{1\over m_b}
\right),  \non  \\
F_2^\bc(\omega) &=& G_2^\bc(\omega)=-{\bar{\Lambda}\over m_c}\,{1\over
1+\omega}, \\
F_3^\bc(\omega) &=& -G_3^\bc(\omega)=-{\bar{\Lambda}\over 2m_b}\,{1\over
1+\omega},  \non \\
G_1^\bc(\omega) &=& 1-{\bar{\Lambda}\over 2}\left({1\over m_c}+{1\over m_b}
\right){1-\omega \over 1+\omega}.   \non
\en
We see that the nonrelativistic quark model predictions for $\Lambda_b\to
\Lambda_c$ form factors at the symmetric point $\omega=1$ are in agreement
with HQET up to the order of $1/m_b$ and $1/m_c$, as it should be.

   In the heavy quark limit, the sextet $\Sigma_b\to\Sigma_c$ or $\Omega_b
\to\Omega_c$ transition at $v\cdot v'=1$ is predicted by HQET to be (see e.g.
Ref.[1])
\be
\la\Sigma_c(v',s')|V_\mu-A_\mu|\Sigma_b(v,s)\ra=\,\bar{u}_{\Sigma_c}(v',s')
[\gamma_\mu(1+\gamma_5)-2(v+v')_\mu]u_{\Sigma_b}(v,s).
\en
This leads to the sextet-sextet baryonic form factors
\be
F_1^{\Sigma_b\Sigma_c}(1)=G_1^{\Sigma_b\Sigma_c}(1)=-{1\over 3},~~
F_2^{\Sigma_b\Sigma_c}(1)=F_3^{\Sigma_b\Sigma_c}(1)=\,{2\over 3},~~
G_2^{\Sigma_b\Sigma_c}(1)=G_3^{\Sigma_b\Sigma_c}(1)=0,
\en
and
\be
f_1^{\Sigma_b\Sigma_c}(q_m^2) &=& -{1\over 3}\left[1-(m_{\Sigma_b}+m_{\Sigma
_c})\left({1\over m_{\Sigma_b}}+{1\over m_{\Sigma_c}}\right)\right],  \non \\
f_2^{\Sigma_b\Sigma_c}(q_m^2) &=& {1\over 3}\left({1\over m_{\Sigma_b}}+
{1\over m_{\Sigma_c}}\right),  \non \\
f_3^{\Sigma_b\Sigma_c}(q_m^2) &=& {1\over 3}\left({1\over m_{\Sigma_b}}-
{1\over m_{\Sigma_c}}\right),   \\
g_1^{\Sigma_b\Sigma_c}(q^2_m) &=& -{1\over 3},~~~~g_2^{\Sigma_b\Sigma_c}
(q^2_m)=g_3^{\Sigma_b\Sigma_c}(q^2_m)=0,  \non
\en
and similar results for $\Omega_b\to\Omega_c$. Since $N_{fi}=1$ and
$\eta=-1/3$ for sextet-sextet heavy baryon transitions, it is evident from
Eqs.(22) and (23) that our quark model results for $\Sigma_b\to\Sigma_c$
and $\Omega_b\to\Omega_c$ form factors evaluated at zero recoil in the heavy
quark limit are in agreement with the constraints imposed by heavy quark
symmetry.

    We now make a comparsion with the quark model calculations in Refs.[5,6].
Quark- and bag-model wave functions in the coordinate space are used to
evaluate the baryonic form factors by P\'erez-Marcial {\it et al.} [5].
However, their results (11a-11f) at zero recoil,
\footnote{The parameters $\alpha_1$ and $\alpha_2$ defined in Eq.(12) of
Ref.[5] correspond to our $N_{fi}$ and $N_{fi}'$, respectively.}
when applied to $\Lambda_b\to\Lambda_c$ and $\Sigma_b\to\Sigma_c$,
are in disagreement with HQET.
In fact, the heavy quark symmetry relations $f_1=g_1,~f_2=g_3$ and $f_3=g_2$
for $\Lambda_b\to\Lambda_c$ transition and $g_1=-1/3$, $g_2=g_3=0$ for
$\Sigma_b\to\Sigma_c$ implied by HQET are not respected
by Ref.[5]. Moreover, the dimensionless $\Lambda_b\to\Lambda_c$
form factors $m_{\Lambda_b}f_{2,3}$ and $m_{\Lambda_b}g_{2,3}$ vanish in
the heavy quark limit according to HQET.
The reader can check that the form factors obtained
in Ref.[5] do not satisfy this feature of heavy quark symmetry.

   Our evaluation of baryonic form factors is quite close to that of
Singleton [6] except mainly for Eq.(16), in which we have
taken the average of the integrations over $(\vec{p}_{12},\vec{\ell}\,)$
and ($\vec{p'}_{12},\vec{\ell}')$ (see the footnote there).
Besides the $1/m_q$ corrections we have also included $1/m_Q$ effects, which
are not taken into account in [6].
Recasting (3.48)-(3.51) of Ref.[6] into the form
factors used here gives
\be
&& f_1^\bc(q^2_m)=g_1^\bc(q^2_m) =\, 1+{\Delta m\over 2m_{\Lambda_b}}\,
{\bar{\Lambda}\over m_c},   \non \\
&& f_2^\bc(q^2_m)=f_3^\bc(q^2_m)= g_2^\bc(q^2_m)=g_3^\bc(q^2_m) =\,-{\bar{
\Lambda}\over 2m_c}\,{1\over m_{\Lambda_b}},
\en
for $\Lambda_b\to\Lambda_c$ transition.
Comparing with (25), it is evident that the $1/m_c$ corrections in (30) are too
large by a factor of 2; that is, the quark model calculations by Singleton
do satisfy
the aforementioned heavy quark symmetry relations, but are still not
consistent with HQET in magnitude.

    The $B_i\to B_f$ baryonic form factors (22) at maximum $q^2$ obtained in
the nonrelativistic quark model are the main results in the present paper.
The $1/m_Q$ and $1/m_q$ effects in (22) arise from the modification to
the current operator. Although as far as $\Lambda_Q\to\Lambda_{Q'}$ and
$\Xi_Q\to\Xi_{Q'}$ are concerned, the nonrelativistic quark model predictions
for the form factors at $\omega=1$ are in accordance with HQET, the two
approaches differ in two main aspects: (i) Unlike the nonrelativistic quark
model, HQET provides a systematic $\Lambda_{\rm QCD}/m_Q$ expansion, which
can be treated perturbatively if $m_Q>>\Lambda_{\rm QCD}$.
Near zero recoil in the rest frame of
the parent baryon, the quark model result for $1/m_q$ corrections is
trustworthy since $|\vec{q}|/m_q<<1$, where $-\vec{q}$ is the
recoil momentum of the daughter baryon. Consequently, contrary to HQET,
$1/m_s$ modifications to the form factors near $v\cdot v'=1$ become meaningful
in the quark model. (ii) Going beyond the antitriplet-antitriplet heavy baryon
transition, HQET loses its predictive power for form factors at order $1/m_Q$
since $1/m_Q$ corrections due to wave function modifications arising
from $O_1$ and $O_2$ are not calculable by perturbative QCD.
However, such corrections are expected to vanish at zero recoil in the
quark model.
\footnote{This is known to be true in the meson case. Among the four
subleading Isgur-Wise functions $\eta(\omega),~\chi_{1,2,3}(\omega)$ (see
the Introduction), we know that $\chi_1$ and $\chi_3$ vanish at $\omega=1$
and that $\eta(\omega)=\chi_2(\omega)=0$ in the quark model [3].}
This is so because the physical results at the symmetric point $\omega=1$,
where both parent and daughter baryons are at rest, should be independent of
the explicit form of the wave function. Modifications to the wave function
come from the operators $O_1$ and $O_2$ acting on $\psi_0$ [see Eq.(1)],
whose explicit expression is model dependent. As a result, the
nonrelativistic quark model results (22) for weak current-induced form factors
evaluated at maximum $q^2$ are applicable to
any heavy-heavy and heavy-light baryonic transitions.

   At this point, we would like to examine the underlying assumptions we have
made during the course of deriving (18) or (22). The assumptions are (i)
the approximation of the weak binding mass $\tilde{m}_f(=m_q+m_1+m_2)$ with
the mass $m_f$ of the daughter baryon, (ii) flavor indpendence of the momentum
wave function, $\phi_f=\phi_i$, and (iii) the average of two momenta
integrals in (16). Assumption (i) is justified by the fact that $\tilde{m}_f$
and $m_f$ for charmed and octet baryons are very close for $m_u=338$ MeV,
$m_d=322$ MeV, $m_s=510$ MeV (see p.1729 of Ref.[7]) and $m_c=1.6$ GeV.
In contrast, assumptions (ii) and (iii) are less solid. First, the momentum
wave function $\phi$ is truly flavor independent only in the heavy
quark limit. Second, a simple average of two different momenta integrals
taken in (16) seems somewhat arbitrary. Nevertheless, the present
prescription works empirically at it does agree with HQET at order $1/m_Q$.
Since the hadronic matrix element (14) should in principle
be independent of the integration order, this probably means that
flavor dependence of $\phi$, which is of order $1/m_Q$ and $1/m_q$ is
compensated by similar effects in (iii). A full understanding of the empirical
agreement between present approach and HQET needs to be pursued.

    Experimentally, the only information available so far is the form-factor
ratio measured in the semileptonic decay $\Lambda_c\to\Lambda e\bar{\nu}$.
In the heavy
$c$-quark limit, there are two independent form factors in $\Lambda_c
\to\Lambda$ transition [8]
\be
\la\Lambda(p)|\bar{s}\gamma_\mu(1-\gamma_5)c|\Lambda_c(v)\ra=\,\bar{u}
_{_\Lambda}\left(F_1^{\Lambda_c\Lambda}(v\cdot p)+v\!\!\!/ F_2^{\Lambda_c
\Lambda}(v\cdot p)\right)\gamma_\mu(1-\gamma_5)u_{_{\Lambda_c}}.
\en
Assuming a dipole $q^2$ behavior for form factors, the ratio
$R=\tilde{F}_1^{\Lambda_c\Lambda}/\tilde{F}_2^{\Lambda_c\Lambda}$ is
measured by CLEO to be [9]
\be
R=-0.25\pm 0.14\pm 0.08\,.
\en
The form factors $\tilde{F}_{1,2}$ are related to $f$'s and $g$'s by
\be
f_1=g_1=\tilde{F}_1+{m_f\over m_i}\tilde{F}_2,~~~f_2=f_3=g_2=g_3={\tilde{F}_2
\over m_i}.
\en
Since $R$ is independent of $q^2$ if $\tilde{F}_1$ and $\tilde{F}_2$ have the
same $q^2$ dependence, we can apply the quark model (22) to get
\be
\tilde{F}_1(q_m^2)=\,1+{\bar{\Lambda}\over 4m_s},~~~~\tilde{F}_2(q^2_m)=
-{\bar{\Lambda}\over 4m_s},
\en
which lead to
\be
R=-\left(1+{4m_s\over\bar{\Lambda}}\right)^{-1}=-0.23
\en
for $m_s=510$ MeV. This is in excellent agreement
with experiment (32), but it should be stressed that $1/m_c$
corrections, which are potentially important, have not been included in (32)
and (35).

\vskip 0.15cm
\noindent{\bf III.~~Applications}
\vskip 0.15cm
    In this section we will apply the baryonic form factors obtained in
the nonrelativistic quark model to various physical processes. Since the
calculation of the $q^2$ dependence of form factors is beyond the scope of
a nonrelativistic quark model, we will thus assume a pole
dominance for the form-factor $q^2$ behavior:
\be
f(q^2)={f(0)\over \left(1-{q^2\over m_V^2}\right)^n}\,,~~~~g(q^2)={g(0)\over
\left(1-{q^2\over m_A^2}\right)^n}\,,
\en
where $m_V$ ($m_A$) is the pole mass of the vector (axial-vector) meson
with the same quantum number as the current under consideration. In practice,
either monopole ($n=1$) or dipole
($n=2$) $q^2$ dependence are adopted in the literature. For definiteness, we
will choose the dipole behavior suggested by the following argument.
Considering the function
\be
G(q^2)=\left( {1-q_m^2/m_*^2\over 1-q^2/m_*^2}\right)^n,
\en
with $m_*$ being the pole mass, it is clear that $G(q^2)$
plays the role of the baryonic Isgur-Wise function $\zeta(v\cdot v')$ in
$\Lambda_Q\to\Lambda_{Q'}$ transition, namely $G=1$ at $q^2=q_m^2$.
The function $\zeta(\omega)$ has been calculated in two different models:
\be
\zeta(\omega)=\cases{ 0.99\exp[-1.3(\omega-1)], & soliton~model~[10]; \cr
\left({2\over \omega+1}\right)^{3.5+{1.2\over\omega}}, & MIT~bag~model~[11].
\cr}
\en
Using the pole masses $m_V=6.34$ GeV and $m_A=6.73$ GeV for the transition
$\Lambda_b\to\Lambda_c$, we find that $G(q^2)$ is compatible with
$\zeta(\omega)$ only if $n=2$. However, one should bear in mind that, in
reality, the $q^2$ behavior of form factors is probably more complicated
and it is likely that a simple pole dominance only applies to a certain
$q^2$ region.

    Before proceeding to applications we would like to make a remark on the
role played by heavy quark symmetry here. Though HQET is employed in Sec. II
as a benchmark for testing if the nonrelativistic quark model calculations
of form factors are reliable and trustworthy,
heavy quark symmetry is no longer relevant in
all applications described in this section except for in Eq.(65) where we
apply the static heavy quark limit to relate the tensor matrix element to
the vector and axial-vector one. For example, the different but realistic
vector and axial-vector pole masses used in (36) break heavy quark symmetry
explicitly.

\vskip 0.12cm
\noindent{\bf 3.1~~Semileptonic decay}
\vskip 0.07cm
     We shall study in this subsection the decay rate for the semileptonic
transition ${1\over 2}^+\to{1\over 2}^++e+\bar{\nu}_e$. Take the semileptonic
decay $\Lambda_c^+\to\Lambda e^+\nu_e$ as an example. Since $\eta=1,~m_V=
m_{D_s(1^-)}=2.11$ GeV, $m_A=m_{D_s(1^+)}=2.536$ GeV [7], the form factors
at $q^2=0$ obtained from (22) and (36) with $n=2$ are
\be
f_1^\cs(0)=\,0.50N_\cs, && f_2^\cs(0)=-0.25N_\cs/m_{\Lambda_c},~~~f_3^
\cs(0)=-0.05N_\cs/m_{\Lambda_c},    \non \\
g_1^\cs(0)=\,0.65N_\cs, && g_2^\cs(0)=-0.06N_\cs/m_{\Lambda_c},~~~g_3^
\cs(0)=-0.32N_\cs/m_{\Lambda_c},
\en
where uses of $m_c=1.6$ GeV and $m_s=510$ MeV have been made.
{}From the flavor-spin wave function of $\Lambda_c$ and $\Lambda$ with a
positive helicity along the $z$ direction
\be
|\Lambda_c\up\ra_{\rm flavor-spin} &=& {1\over\sqrt{2}}(ud-du)c\chi_A,   \non
\\
|\Lambda\up\ra_{\rm flavor-spin} &=& {1\over\sqrt{6}}[(ud-du)s\chi_A+(13)
+(23)],
\en
where $(ij)$ means permutation for the quark in place $i$ with the quark
in place $j$, we get
\be
N_\cs=\,_{\rm flavor-spin}\la \Lambda\up|b_s^\dagger b_c|\Lambda_c\up\ra_{\rm
flavor-spin}=\,{1\over\sqrt{3}}.
\en
The computation of the baryon semileptonic decay rate is straightforward;
for an analytic expression of the decay rate, see for example Ref.[12].
We obtain
\be
\Gamma(\Lambda_c\to\Lambda e^+\nu_e)=\,(N_\cs)^2\times 2.11\times 10^{11}s^
{-1}=\,7.1\times 10^{10}s^{-1},
\en
which is in excellent agreement with experiment [7]
\be
\Gamma(\Lambda_c\to\Lambda e^+\nu_e)_{\rm expt}=\,(7.0\pm 2.5)\times
10^{10}s^{-1}.
\en

\vskip 0.5cm
{\small Table I. Nonrelativistic quark model predictions for
baryonic form factors evaluated at $q^2=0$ using dipole $q^2$ dependence
and $|V_{cb}|=0.040$ [15] ($m_i$ being the mass of the parent
heavy baryon). For a comparsion, we also present the nonrelativistic quark
model predictions given in Refs.[5] and [6]. The numerical values of the
former reference are quoted
from Table IV of [5], while the predicted values of the latter are
computed from Eqs.(3.48)-(3.51) of [6]. Pole masses are taken to be
$m_V=2.11$ GeV, $m_A=2.536$ GeV for $B_c\to B$ transition and $m_V=6.34$ GeV,
$m_A=6.73$ GeV for $B_b\to B_c$. Also shown are the spin and flavor factors
for various baryonic transitions.}
\begin{center}
\begin{tabular}{|c||c|c||c|c|c|c|c|c||c|} \hline
transition & $\eta$ & $N_{fi}$ & $f_1(0)$ & $f_2(0)m_i$ & $f_3(0)m_i$ &
$g_1(0)$ & $g_2(0)m_i$ & $g_3(0)m_i$ &  \\  \hline
$\Lambda_c^+\to\Lambda^0$ & 1 & ${1\over\sqrt{3}}$ & 0.29 & $-0.14$ & $-0.03$
& 0.38 & $-0.03$ & $-0.19$ & this work   \\
& 1 & 1 & 0.35 & $-0.09$ & 0.25 & 0.61 & $0.04$ & $-0.10$ & [5]   \\
& 1 & ${1\over\sqrt{3}}$ & 0.36 & $-0.17$ & $-0.17$ & 0.47 & $-0.22$ & $-0.22$
& [6]   \\ \hline
$\Xi_c^0\to \Xi^-$ & 1 & ${1\over\sqrt{3}}$ & 0.31 & $-0.19$ & $-0.04$ & 0.39
& $-0.06$ & $-0.24$ & this work   \\
& 1 & 1 & 0.48 & $-0.08$ & 0.26 & 0.76 & $0.04$ & $-0.12$ & [5]   \\
& 1 & ${1\over\sqrt{3}}$ & 0.40 & $-0.23$ & $-0.23$ & 0.50 & $-0.30$ & $-0.30$
& [6]   \\ \hline
$\Lambda_b^0\to\Lambda_c^+$ & 1 & 1 & 0.53 & $-0.12$ & $-0.02$ & 0.58 &
$-0.02$ & $-0.13$ & this work   \\
& 1 & 1 & 0.60 & $-0.14$ & $-0.14$ & 0.63 & $-0.15$ & $-0.15$ & [6]\\ \hline
$\Xi_b^0\to \Xi_c^+$ & 1 & 1 & 0.54 & $-0.14$ & $-0.02$ & 0.58 & $-0.03$ &
$-0.16$ & this work   \\
& 1 & 1 & 0.62 & $-0.17$ & $-0.17$ & 0.67 & $-0.18$ & $-0.18$ & [6] \\ \hline
$\Omega_b^-\to\Omega_c^0$ & $-{1\over 3}$ & 1 & 0.72 & 0.68 & $-0.36$ &
$-0.20$ & 0.01 & 0.06 & this work   \\
& $-{1\over 3}$ & 1 & 0.85 & 0.81 & $-0.60$ & $-0.23$ & 0.08 & 0.08 & [6] \\
\hline
\end{tabular}
\end{center}

    It must be stressed that the flavor factor $N_\cs=1/\sqrt{3}$, which was
already noticed in [6,13] and particularly accentuated by [14], is very
crucial for an agreement between theory
and experiment. In the literature it is customary to replace the $s$ quark
in the baryon $\Lambda$ by the heavy quark $Q$ to obtain the wave function
of the $\Lambda_Q$. However, this amounts to assuming SU(4) or SU(5) flavor
symmetry. Since SU(N)-flavor symmetry with $N>3$
is badly broken, the flavor factor $N_{\Lambda_Q\Lambda}$ is no longer unity
(of course, $N_{\Lambda_Q\Lambda_{Q'}}=1$). Indeed, if $N_\cs$ were equal to
one, the predicted rate for $\Lambda_c\to\Lambda e^+\nu_e$ would have been
too large by a factor of 3 !

    For completeness, the numerical values of form factors at $q^2=0$
are tabulated in Table I and the nonrelativistic quark model predictions
for the decay rates of semileptonic decays of heavy baryons are summarized in
Table II.
We will not consider the case of sextet heavy baryons as they are
dominated by strong or electromagnetic decays (except for $\Omega_Q$).
Two remarks are in order. (i) We see from Tables I and II that the predictions
of Ref.[5] for form factors and semileptonic decay rates are in general
different substantially from ours. First, the important suppression factor of
$N_{fi}$ for antitriplet heavy baryon-octet baryon transition is not
taken into account in [5]. Second, the calculated heavy-heavy baryon
form factors in [5] at zero recoil
do not satisfy the constraints imposed by heavy quark symmetry.
The results of Ref.[6] are
more close to ours. However, $1/m_Q$ corrections are not included in [6] and
the computed $1/m_q$ effects there disagree with HQET for
$\Lambda_b\to\Lambda_c$ and $\Xi_b\to\Xi_c$ transitions.
(ii) The parameter $\bar{\Lambda}$ is
process dependent; for example, it can be as large as 1.11 GeV for $\Omega_b
\to\Omega_c e\bar{\nu}$, whereas it is only 0.61 GeV for $\Lambda_c\to\Lambda
e^+\nu_e$.

\vskip 0.5cm
{\small Table II. Nonrelativistic quark model predictions for
the semileptonic decay rates in units of $10^{10}s^{-1}$ evaluated using
dipole $q^2$ dependence for form factors. Values in parentheses are the
predicted rates with QCD corrections. However, as stressed in [14], it seems
that QCD effects computed in [5] are unrealistically too large.}
\begin{center}
\begin{tabular}{|c||c|c|c||c|} \hline
 process & [5] & [6] & this work & experiment [7] \\ \hline
$\Lambda_c^+\to\Lambda^0 e^+\nu_e$  & 18.0~(11.2) & 9.8 & 7.1
& $7.0\pm 2.5$ \\ \hline
$\Xi_c^0\to\Xi^- e^+\nu_e$  & 28.8~(18.1) & 8.5 & 7.4 & - \\ \hline
$\Lambda_b^0\to\Lambda_c^+ e^-\bar{\nu}_e$  & - & 5.9 & 5.1 & - \\ \hline
$\Xi_b^0\to\Xi_c^+ e^-\bar{\nu}_e$  & - & 7.2 & 5.3 & - \\ \hline
$\Omega_b^-\to\Omega_c^0 e^-\bar{\nu}_e$ & - & 5.4 & 2.3 & - \\ \hline
\end{tabular}
\end{center}

\vskip 0.3cm
\noindent{\bf 3.2~~Nonleptonic decay}
\vskip 0.1cm
At the quark level,
the nonleptonic weak decays of the baryon usually receive contributions
from external $W$-emission, internal $W$-emission and $W$-exchange diagrams.
At the hadronic level, these contributions manifest as factorizable and
pole diagrams. It is known that, contrary to the meson case, the nonspectator
$W$-exchange effects in charmed baryon decays are of comparable importance
as the spectator diagrams [16]. Unfortunately, in general it is difficult to
estimate the pole diagrams. Nevertheless, there exist some decay modes
of heavy baryons which proceed only through the internal or external
$W$-emission diagram. Examples are
\be
 {\rm internal}~W{\rm -emission} &:&
\b,~~\Xi_b\to J/\psi\Xi,~~\Omega_b\to J/\psi\Omega,~~\Lambda_c\to p\phi,
\cdots \non \\
 {\rm external}~W{\rm -emission} &:&
\Omega_b\to\Omega_c\pi,~~~\Omega_c\to\Omega\pi.
\en
Consequently, the above decay modes are free of nonspectator effects and their
theoretical calculations are relatively clean.

    In this subsection we shall study two of the decay modes displayed in
(44), namely $\b$ and $\Lambda_c\to p\phi$.
The general amplitude of $\b$ has the form
\be
A(\b)=i\bar{u}_\Lambda(p_\Lambda)\varepsilon^{*\mu}[A_1\gamma_\mu\gamma_5+
A_2(p_\Lambda)_\mu\gamma_5+B_1\gamma_\mu+B_2(p_\Lambda)_\mu]u_{\Lambda_b}
(p_{\Lambda_b}),
\en
where $\ep_\mu$ is the polarization vector of the $\j$.
Under factorization assumption, the internal $W$-emission contribution reads
\be
A(\b)=\,{G_F\over\sqrt{2}}V_{cb}V_{cs}^*a_2\la J/\psi|\bar{c}\gamma_\mu(1-
\gamma_5)c|0\ra\la\Lambda|\bar{s}\gamma^\mu(1-\gamma_5)b|\Lambda_b\ra,
\en
where $a_2$ is an unknown parameter introduced in Ref.[17]. It follows from
(45) and (46) that
\be
A_1 &=& -\lambda[g_1^\bs(m^2_\j)+g_2^\bs(m^2_\j)(m_{\Lambda_b}-m_\Lambda)],
\non \\
A_2 &=& -2\lambda g_2^\bs(m^2_\j),   \non \\
B_1 &=& \lambda[f_1^\bs(m^2_\j)-f_2^\bs(m^2_\j)(m_{\Lambda_b}+m_\Lambda)],  \\
B_2 &=& 2\lambda f_2^\bs(m^2_\j),   \non
\en
with $\lambda={G_F\over\sqrt{2}}V_{cb}V_{cs}^*a_2f_\j m_\j$. Since $\eta=1$,
$N_{\bs}={1\over\sqrt{3}}$, $m_V=m_{B_s(1^-)}\cong 5.42$ GeV and $m_A=m_{
B_s(1^+)}\cong 5.86$ GeV, we find from Eqs.(22) and (36) that
\be
&& f_1^\bs(m_\j^2)=0.131,~~~~f_2^\bs(m_\j^2)=-0.054/m_{\Lambda_b},   \non \\
&& g_1^\bs(m_\j^2)=0.203,~~~~g_2^\bs(m_\j^2)=-0.036/m_{\Lambda_b}.
\en

The decay rate reads [18]
\be
\Gamma(\b)={p_c\over 8\pi}\,{E_\j+m_\j\over m_{\Lambda_b}}\left[
2(|S|^2+|P_2|^2)+{E^2_\j\over m_\j^2}(|S+D|^2+|P_1|^2)\right],
\en
with the $S,~P$ and $D$ waves given by
\be
S &=& -A_1,   \non \\
P_1 &=& -{p_c\over E_\j}\left({m_{\Lambda_b}+m_\Lambda\over E_\Lambda+m
_\Lambda}B_1+m_{\Lambda_b}B_2\right),    \non \\
P_2 &=& {p_c\over E_\Lambda+m_\Lambda}B_1,   \\
D &=& -{p_c^2\over E_\j(E_\Lambda+m_\Lambda)}\,(A_1-m_{\Lambda_b}A_2), \non
\en
where $p_c$ is the c.m. momentum. Using $|V_{cb}|=0.040$ [15],
$\tau(\Lambda_b)=1.07\times 10^{-12}s$ [7], $a_2\sim 0.23$ [19],
and $f_\j=395\,$MeV extracted from the observed $J/\psi\to e^+e^-$ rate,
$\Gamma(\j\to e^+e^-)=(5.27\pm 0.37)$ keV [7],
we find
\be
{\cal B}(\b)=\,2.1\times 10^{-4}\,.
\en
When
anisotropy in angular distribution is produced in a polarized $\Lambda_b$
decay, it is governed by the asymmetry parameter $\alpha$ given by [18]
\be
\alpha=\,{4m^2_\j{\rm Re}(S^*P_2)+2E^2_\j{\rm Re}(S+D)^*P_1\over 2m_\j^2
(|S|^2+|P_2|^2)+E^2_\j(|S+D|^2+|P_1|^2)}.
\en
Numerically, it reads
\footnote{Our previous study on $\b$ [20] has a vital sign error in
Eq.(14) for the expression of the $D$-wave amplitude, which affects the
magnitude of the decay rate and the sign of decay asymmetry. Moreover, the
important flavor factor $N_{\Lambda_b\Lambda}=1/\sqrt{3}$ is not taken into
account there.}
\be
\alpha(\b)=\,-0.11\,,
\en
where the negative sign of $\alpha$ reflects the $V-A$ structure of
the current.

   The $\b$ decay was originally reported by the
UA1 Collaboration [21] with the result
\be
F(\Lambda_b){\cal B}(\b)=\,(1.8\pm 0.6\pm 0.9)\times 10^{-3},
\en
where $F(\Lambda_b)$ is the fraction of $b$ quarks fragmenting into
$\Lambda_b$. Assuming $F(\Lambda_b)=10\%$ [21], this leads to
\be
{\cal B}(\b)=\,(1.8\pm 1.1)\%.
\en
However, both CDF [22] and LEP [23] did not see any evidence for this decay.
For example, based on the
signal claimed by UA1, CDF should have reconstructed $30\pm 23$ $\b$ events.
Instead CDF found not more than 2 events and concluded that
\be
F(\Lambda_b){\cal B}(\b)< 0.50\times 10^{-3}.
\en
The limit set by OPAL is [23]
\be
F(\Lambda_b){\cal B}(\b)< 1.1\times 10^{-3}.
\en
Hence, a theoretical study of this decay mode
would be quite helpful to clarify the issue. The prediction (51) indicates
that the branching ratio we obtained is two orders of magnitude smaller
than what expected from UA1 (55).

     We next turn to the Cabibbo-suppressed decay $\Lambda_c\to p\phi$. As
emphasized in Ref.[16], this decay mode is of particular interest because
it provides a direct test of the large-$N_c$ approach in the charmed baryon
sector, though this approach is known to
work well for the nonleptonic weak decays of charmed mesons.
{}From the flavor-spin wave function of the
$\Lambda_c$ (40) and the proton
\be
|p\up\ra_{\rm flavor-spin}=\,{1\over\sqrt{3}}[uud\chi_s+(13)+(23)],
\en
we get
\be
N_{\Lambda_c p}=\,{1\over\sqrt{2}}.
\en
Since the calculation is very similar to that of $\b$, we simply write down
the results:
\be
{\cal B}(\Lambda_c\to p\phi)=\,(c_2)^2\,2.26\times 10^{-3},~~~~
\alpha(\Lambda_c\to p\phi)=-0.10\,,
\en
where we have applied $f_\phi=237$ MeV, $m_{D(1^-)}=2.01$ GeV and $m_{D(1^+)}
=2.42$ GeV. As the Wilson coefficient $c_2$ is expected to be of order
$-0.56$ in the large-$N_c$ approach, it follows that
\be
{\cal B}(\Lambda_c\to p\phi)=\,7.1\times 10^{-4}\,.
\en
Therefore, in order to test $1/N_c$ expansion and the nonrelativistic quark
model for the form factors, the experimental accuracy should be reached at
the level of a few $10^{-4}$. Experimentally, the branching ratio is
measured to be
\be
{\cal B}(\Lambda_c\to p\phi)=\cases{ (1.8\pm 1.2)\times 10^{-3}, & ACCMOR~[24];
\cr    <1.7\times 10^{-3}, & E687~[25].   \cr}
\en

Finally, it is worth remarking that it is important to take into account the
effect of the flavor-suppression factor (e.g. $N_{\Lambda_c\Lambda}$) on the
factorizable contributions to the nonleptonic two-body decays of charmed
baryons; such effects thus far have not been considered in the literature
[16,26].

\vskip 0.2cm
\noindent{\bf 3.3~~Weak radiative decay}
\vskip 0.15cm
Recently the weak radiative decays of $B$ mesons and bottom baryons have been
systematically
studied in Ref.[27]. At the quark level, there are two essential mechanisms
responsible for weak radiative decays: electromagnetic penguin mechanism
and $W$-exchange (or $W$-annihilation) bremsstrahlung. The two-body decays of
the bottom baryons
proceeding through the short-distance electromagnetic penguin diagrams are:
\be
\Lambda_b^0\to\Sigma^0\gamma,~\Lambda^0\gamma,~~~~\Xi_b^0\to\Xi^0\gamma,~~~~
\Xi_b^-\to\Xi^-\gamma,~~~~\Omega_b^-\to\Omega^-\gamma.
\en
In this subsection, we shall study the above weak
radiative decay modes using the nonrelativistic quark model in
conjunction with the heavy $b$-quark symmetry.

To begin with, the
electromagnetic penguin-induced radiative decay amplitude is [27]
\be
A(B_i\to B_f+\gamma) &=& i{G_F\over \sqrt{2}}\,{e
\over 8\pi^2}F_2(x_t)V_{tb}V_{ts}^*\,m_b\ep^\mu q^\nu  \non \\
&\times& \la B_f|\bar{s}\sigma_{\mu\nu}[(1+\gamma_5)+{m_s\over m_b}
(1-\gamma_5)]b|B_i\ra,
\en
where $q$ is the photon momentum, $F_2$ is a smooth function of $x_t\equiv
m_t^2/M_W^2$ [28] and it is numerically equal to 0.65 for $\Lambda_{\rm
QCD}=200$ MeV and $m_t=174$ GeV. In order to evaluate the tensor matrix
elements in (64), we consider the static heavy $b$-quark limit so that
\be
\la B_f|\bar{s}i\sigma_{0i}(1+\gamma_5)b|B_i\ra=\,\la B_f|
\bar{s}\gamma_i(1-\gamma_5)b|B_i\ra.
\en
Hence,
\be
 \la B_f|\bar{s}i\sigma_{0i}(1+\gamma_5)b|B_i\ra \ep^0q^i &=& {1\over 2}\la
B_f|\bar{s}\gamma_i(1-\gamma_5)b|B_i\ra(\ep^0q^i-\ep^iq^0)     \\
 &=& \bar{u}_fi\sigma_{0i}\ep^0q^i[f_1-f_2(m_i+m_f)+g_1\gamma_5+g_2(m_i-m_f)
\gamma_5]u_i.     \non
\en
It follows from (64) and (66) that
\be
A(B_i\to B_f+\gamma)=\,i\bar{u}_f(a+b\gamma_5)\sigma_{\mu\nu}\ep^\mu q^\nu
u_i,
\en
with
\be
a &=& {G_F\over \sqrt{2}}\,{e\over 8\pi^2}F_2(x_t)m_bV_{tb}V^*_{ts}\,[f_1(0)
-f_2(0)(m_i+m_f)],   \non \\
b &=& {G_F\over \sqrt{2}}\,{e\over 8\pi^2}F_2(x_t)m_bV_{tb}
V^*_{ts}\,[g_1(0)+g_2(0)(m_i-m_f)],
\en
being parity-conserving and -violating amplitudes, respectively. The decay
rate is
\be
\Gamma(B_i\to B_f+\gamma)=\,{1\over 8\pi}\left( {m^2_i-m^2_f\over m_i}\right)
^3(|a|^2+|b|^2).
\en

    In order to apply the heavy quark symmetry relation (65), we shall
neglect $1/m_b$ corrections to the form factors given in (22).
To the leading order in $1/m_b$, we obtain
\be
\Gamma(\Lambda_b\to\Lambda\gamma) &=& 1.6\times 10^{-18}\,{\rm GeV}, \non \\
\Gamma(\Xi_b\to\Xi\gamma) &=& 2.2\times 10^{-18}\,{\rm GeV},
\en
and a prohibited $\Lambda_b\to\Sigma\gamma$, where uses of
$N_{\Lambda_b\Lambda}=N_{\Xi_b\Xi}=1/\sqrt{3}$ and
$V_{tb}V_{ts}^*\approx -V_{cb}V_{cs}^*$ have been made. Therefore,
\be
{\cal B}(\Lambda_b\to\Lambda\gamma)=2.7\times 10^{-6},
\en
for $\tau(\Lambda_b)=1.07\times 10^{-12}s$ [7]. In Ref.[27] two different
methods, namely the heavy $s$-quark approach and the MIT bag model, have
been employed to estimate the decay rate of $\Lambda_b\to\Lambda\gamma$.
Our present result (71) is somewhat smaller than the prediction given in [27]
owing to the presence of the flavor-suppression factor of $1/\sqrt{3}$
in the amplitude.

\vskip 0.15cm
\noindent{\bf IV.~~Conclusions and discussion}
\vskip 0.15cm
    In the heavy quark effect theory (HQET),
    current-induced $1/m_Q$ corrections and the presence of higher dimensional
operators in the effective Lagrangian are the two sources of $1/m_Q$
effects on the hadronic form factors. Since the predictive power of HQET for
baryon form factors at order $1/m_Q$ is limited only to antitriplet-antitriplet
heavy baryon transition, this motivates us to apply
the nonrelativistic quark model to evaluate the weak current-induced baryonic
form factors at zero recoil in the rest frame of the heavy parent baryon,
where the quark model is most trustworthy. Contrary to previous similar
work, we have shown that the HQET predictions for
antitriplet-antipriplet heavy baryon transitions at $v\cdot v'=1$
are reproducible in the nonrelativistic quark model. Moreover, the latter
approach has two eminent features. First, it becomes meaningful
to consider $1/m_s$ corrections so long as the recoil momentum is smaller
than the $m_s$ scale. Second, $1/m_Q$ effects arising from wave-function
modifications vanish at zero recoil in the quark model. Consequently, the
nonrelativistic quark model results for the form factors evaluated at
maximum $q^2$ are applicable to any heavy-heavy and heavy-light baryonic
transitions.

  An obvious criterion for testing the reliability of quark model calculations
is that model results must satisfy all the constraints imposed by heavy quark
symmetry. In the heavy quark limit, normalizations of heavy-heavy form factors
and hence some relations between form factors at zero recoil are fixed by
heavy quark symmetry. These constraints are not respected in Ref.[5]. While
this discrepancy is improved in the work of [6], its prediction for $\Lambda
_b\to\Lambda_c$ (or $\Xi_b\to\Xi_c$) form factors at order $1/m_Q$ is still
too large by a factor of 2 when compared with HQET. We have shown that our
prescription of quark model calculations does incorporate the features
of heavy quark symmetry (a careful examination on the underlying assumptions
we have made is discussed in Sec.II) and hence can be applied to compute
baryon form factors beyond the arena of HQET.

    As the conventional practice, we make the pole dominance assumption for
the $q^2$ dependence to extrapolate the form factors from maximum $q^2$ to
the desired $q^2$ point. We argued that a dipole $q^2$ behavior is more
preferred since it is close to the baryonic Isgur-Wise function calculated
recently. Nevertheless, one should bear in mind that the assumption of
pole dominance for form factors is probably too simplified and this problem
remains unresolved.

     We have applied our main results (22) in this paper to various
decays of heavy baryons. In all model applications described in Sec.III,
heavy quark symmetry is no longer relevant except for in Eq.(65).
It turns out that the inclusion of
a flavor suppression factor, which occurs in most of heavy to light baryonic
transitions, is very crucial to explain the experimental observation of the
semileptonic decay $\Lambda_c\to\Lambda e^+\nu_e$. The presence of this
flavor suppression factor, which is missed in most literature,
will of course affect the predictions on the decay rates of
many decay modes involving a transition from heavy to light baryons. It is
conceivable that some of our predictions can be tested soon in the near future.
Examples are $\Xi_c^0\to\Xi^- e^+\nu_e,~\Lambda_b\to\Lambda e^-\bar{\nu}_e$,
$\b,~\Lambda_c\to p\phi$ and $\Lambda_b\to\Lambda\gamma$. A particularly
interesting decay mode is the channel $\Lambda_b\to\j\Lambda$. Its
branching ratio is predicted to be $2\times 10^{-4}$, which is two orders
of magnitude smaller than the UA1 observation but consistent with the
limit set by CDF and LEP.

\vskip 2.0 cm
\centerline{\bf ACKNOWLEDGMENT}
\vskip 0.5cm
    This work was supported in part by the National Science Council of ROC
under Contract No. NSC84-2112-M-001-014.

\pagebreak
\vskip 1.1 cm
\centerline{\bf REFERENCES}
\vskip 0.3 cm
\begin{enumerate}

\item For a review, see M. Neubert, {\sl Phys. Rep.} {\bf 245}, 261 (1994).

\item M. Luke, \pl {\bf B252}, 447 (1990).

\item J.F. Amundson, \pr {\bf D49}, 373 (1994).

\item H. Georgi, B. Grinstein, and M.B. Wise, \pl {\bf B252}, 456 (1990).

\item R. P\'erez-Marcial, R. Huerta, A. Garcia, and M. Avila-Aoki, \pr
{\bf D40}, 2955 (1989); {\sl ibid.} {\bf D44}, 2203(E) (1991).

\item R. Singleton, \pr {\bf D43}, 2939 (1991).

\item Particle Data Group, \pr {\bf D50}, 1173 (1994).

\item T. Mannel, W. Roberts, and Z. Ryzak, \np {\bf B355}, 38 (1991); \pl
{\bf B255}, 593 (1991); F. Hussain, J.G. K\"orner, M. Kr\"amer, and G.
Thompson, \zp {\bf C51}, 321 (1991).

\item CLEO Collaboration, G. Crawford {\it et al.,} CLNS 94/1306, CLEO 94-24.

\item E. Jenkins, A. Manohar, and M.B. Wise, \np {\bf B396}, 38 (1993).

\item M. Sadzikowski and K. Zalewski, {\sl Z. Phys.} {\bf C59}, 677 (1993).

\item A. Garcia, B. Gonz\'alez, and R. Huerta, \pr {\bf D37}, 2537 (1988).

\item P. Kroll, B. Quadder, and W. Schweiger, \np {\bf B316}, 373 (1989).

\item J.G. K\"orner, M. Kr\"amer, and D. Pirjol, DESY 94-095, hep-ph/9406359.

\item M. Neubert, \pl {\bf B338}, 84 (1994).

\item See e.g. H.Y. Cheng and B. Tseng, \pr {\bf D46}, 1042 (1992); {\sl
ibid.} {\bf D48}, 4188 (1993).

\item M. Bauer, B. Stech, and M. Wirbel, \zp {\bf C34}, 103 (1987).

\item See e.g. S. Pakvasa, S.F. Tuan, and S.P. Rosen, \pr {\bf D42}, 3746
(1990).

\item H.Y. Cheng and B. Tseng, \pr {\bf D51}, 6259 (1995).

\item H.Y. Cheng and B. Tseng, IP-ASTP-22-94, hep-ph/9411215.

\item UA1 Collaboration, C. Albarjar {\it et al.,} \pl {\bf B273}, 540
(1991).

\item CDF Collaboration, F. Abe {\it et al.,} \pr {\bf D47}, 2639 (1993).

\item S.E. Tzmarias, invited talk presented in the 27th International
Conference on High Energy Physics, Glasgow, July 20-27, 1994.

\item ACCMOR Collaboration, S. Barlag, \zp {\bf C48}, 29 (1990).

\item E687 Collaboration, P.L. Farbetti, \pl {\bf B314}, 477 (1993).

\item Q.P. Xu and A.N. Kamal, \pr {\bf D46}, 270 (1992).

\item H.Y. Cheng, C.Y. Cheung, G.L. Lin, Y.C. Lin, T.M. Yan, and H.L. Yu,
\pr {\bf D51}, 1199 (1995).

\item M. Misiak, \pl {\bf B269}, 161 (1991); \np {\bf B393}, 23 (1993); G.
Cella, G. Curci, G. Ricciardi, and A. Vicer\'e, \pl {\bf B248}, 181 (1990);
{\sl ibid.} {\bf B325}, 227 (1994); B. Grinstein, R. Springer, and M.B. Wise,
\np {\bf B339}, 269 (1990); R. Grigjanis, P.J. O'Donnell, M.
Sutherland, and H. Navelet, \pl {\bf B213}, 355 (1988); {\sl ibid.}
{\bf B286}, 413(E) (1992); {\sl Phys. Rep.} {\bf 228}, 93 (1993).

\end{enumerate}

\end{document}